%% file: main.tex
\documentclass[chi]{sigchi}

\makeatletter
\def\@copyrightspace{\relax}
\makeatother

\usepackage{balance}  
\usepackage{graphics} 
\usepackage[T1]{fontenc}
\usepackage{txfonts}
\usepackage{mathptmx}
\usepackage[pdftex]{hyperref}
\usepackage[textsize=tiny]{todonotes}
\usepackage{microtype} 
\usepackage{ccicons}  
\usepackage{etoolbox}
\usepackage{tabu}

\AtBeginEnvironment{quote}{\singlespacing\small \vspace{-16pt}}
\AtEndEnvironment{quote}{\vspace{-6pt}}

\def\plaintitle{Characterizing Scalability Issues in \\ Spreadsheet Software using Online Forums}

\def\emptyauthor{} 
\def\plainkeywords{Reddit; Scalability; Spreadsheets; Excel}

\makeatletter
\def\url@leostyle{%
  \@ifundefined{selectfont}{
    \def\UrlFont{\sf}
  }{
    \def\UrlFont{\small\bf\ttfamily}
  }}
\makeatother

    {\end{list}}

\newcommand{\squishlist}{
   \begin{list}{$\bullet$}
    { \setlength{\itemsep}{0pt}
      \setlength{\parsep}{2pt}
      \setlength{\topsep}{0pt}
      \setlength{\partopsep}{0pt}
      \leftmargin=25pt
\rightmargin=0pt
\labelsep=5pt
\labelwidth=10pt
\itemindent=0pt
\listparindent=0pt
\itemsep=\parsep
    }
}
\newcommand{\squishend}{\end{list}}

\newcommand{\eat}[1]{}
\newcommand{\papertext}[1]{}
\newcommand{\techreport}[1]{#1}

\newcommand{\annon}[1]{}

\def\pprw{8.5in}
\def\pprh{11in}

\definecolor{linkColor}{RGB}{6,125,233}
\hypersetup{%
  pdftitle={\plaintitle},
  pdfauthor={\emptyauthor},
  pdfkeywords={\plainkeywords},
  bookmarksnumbered,
  pdfstartview={FitH},
  colorlinks,
  citecolor=black,
  filecolor=black,
  linkcolor=black,
  urlcolor=linkColor,
  breaklinks=true}

\newcommand{\cut}[1]{}

\begin{document}

\title{\plaintitle}

\numberofauthors{1}
\author{%
  \alignauthor{Kelly Mack, John Lee, Kevin Chang,\\ Karrie Karahalios, Aditya Parameswaran}\\
  \affaddr{Department of Computer Science}\\
    \affaddr{University of Illinois, Urbana-Champaign}\\
  \email{knmack2, lee98, kcchang, kkarhal, adityagp@illinois.edu}\\
}

\maketitle

\begin{abstract}
  In traditional usability studies, researchers  talk to users of tools to understand their needs and challenges. Insights gained via such interviews offer context, detail, and background. Due to costs in time and money, we are beginning to see a new form of tool interrogation that prioritizes scale, cost, and breadth by utilizing existing data from online forums. In this case study, we set out to apply this method of using online forum data to a specific issue---challenges that users face with Excel spreadsheets. Spreadsheets are a versatile and powerful processing tool if used properly. However, with versatility and power come errors, from both users and the software, which make using spreadsheets less effective. By scraping posts from the website Reddit, we collected a dataset of questions and complaints about Excel. Specifically, we explored and characterized the issues users were facing with spreadsheet software in general, and in particular, as resulting from a large amount of data in their spreadsheets. We discuss the implications of our findings on the design of next-generation spreadsheet software.
  \raggedbottom
\end{abstract}
\input{00-intro}
\input{01-introduction}
\input{01.5-reddit}
\input{02-relatedwork}
\input{03-methodology}
\input{04-findings}
\input{05-evaluation}
\input{06-discussions}
\input{07-conclusion}

\bibliographystyle{SIGCHI-Reference-Format}
\bibliography{reference}
\end{document}

%% file: 00-intro.tex
\section{Introduction}
Spreadsheets are widely used by professionals in a variety of disciplines. Even in their most basic form, they act as a useful organizational aid, allowing users to record information in a two-dimensional tabular layout. With capabilities like formulae and macros to support complex calculations or automate processes, spreadsheets become indispensable as a comprehensive  medium for data management and analysis \cite{reschenhofer2015empirical}. 

At the same time, spreadsheets have problems: studies have shown that they foster errors \cite{rajalingham2008classification,panko2016we}, they are not able to easily express useful data operations, such as joins or filters \cite{bakke2016expressive,bakke2011spreadsheet}, and they are known to be sluggish or crash when operating on very large datasets \cite{reschenhofer2015empirical}. The last class of problems are what we call {\em scalability} problems, i.e., situations where users are unable to complete standard operations in a reasonable amount of time when the scale or complexity of the dataset increases \cite{alexander2014microsoft}. With the increasing availability of large volumes of data in most domains, we suspect that scalability problems are only going to become more common---as an example, our biologist collaborators routinely use spreadsheets, and can now generate large genomic (VCF) files, but are unable to open these files even simply to verify correctness in spreadsheet software \cite{bendre2017towards}.

{\em We conduct a case study of problems in spreadsheet software, with a special emphasis on scalability problems.} Prior studies on spreadsheet software has largely focused on the ability of users to use a spreadsheet effectively, as opposed to spreadsheet software problems, including understanding users' conceptual models \cite{mittermeir2002finding, panko2016we, nardi1990ethnographic}, and detecting and categorizing user errors \cite{aurigemma2014evaluating, kruck2001spreadsheet, powell2008critical,chambers2010struggling}. In contrast, we focus on settings where spreadsheet software does not meet expectations. For example, one user performed a copy/paste operation on a few hundred thousand rows of data; they waited two and a half hours and still could not accomplish their task. This behavior---where expectations and models of use are clear---is relatively unexplored.

A second difference with prior studies on understanding the use of spreadsheet software is our methodology. Our first attempt at characterizing spreadsheet problems was via interviews with current or aspiring spreadsheet users, such as our biologist collaborators, teaching faculty who managed MOOC student grades in spreadsheets, business analysts wanting to use spreadsheets to study ad campaigns, among others. While these interviews were illuminating in their depth of detail, they were lacking breadth---we wanted to survey a broad population of spreadsheet users, specifically Microsoft Excel users, to understand their problems. Thus, we turned to online forums, specifically Reddit, that captured users' self-reported opinions, questions, and frustrations on the Microsoft Excel spreadsheet subreddit\footnote{http://reddit.com/r/excel}.

To the best of our knowledge, online forums have not been used to understand the spectrum of problems in spreadsheet software. Online forums have been used for other goals, such as understanding the factors that (1) influence perception---e.g., the discovery of credibility factors in medical crowdfunding sites using Reddit \cite{kim2016power}, (2) describe use and behavior practices---e.g., physical interaction of users with disabilities  using YouTube videos \cite{anthony2013analyzing}, and (3) highlight areas for improvement via ``app mining''---prioritizing application bug fixes using Google Play Store reviews \cite{KeertipatiSL16}. While our work is similar in its categorization of improvement areas to \cite{KeertipatiSL16}, they focus on improving an existing code/review pipeline to determine where to modify code, whereas we investigate the scalability landscape surrounding multiple existing use cases in spreadsheets.  

To accomplish this, we collected over 700 posts from an Excel Reddit forum. While analyzing the posts, four main types of operations emerged that posed problems, both in scalability, and otherwise---importing, managing, querying, and presenting data. We dig deeper to characterize these problem areas to highlight concrete areas of improvement for spreadsheet software, with an eye towards expanding the reach and usability of spreadsheets, especially for very large and complex datasets.

The contributions of this paper are (1) a mapping of challenges users face using spreadsheets in general, as well as (2) how they pertain to scalability, and (3) a methodology for broad evaluation of problems in features, capabilities, and intent specification for end-user software, and (4) a discussion of how to fix these problems, as a means towards building a more robust, scalable, powerful spreadsheet tool. 

%% file: 01-introduction.tex
\section{Related Work}

Our work builds on prior work in two different research areas, (1) spreadsheets and (2) the use of online community discourse as a primary data source for study.

\subsection{Prior work in spreadsheets}
Previous research in spreadsheets have focused on improving spreadsheets through understanding existing problems via user studies and analyzing existing spreadsheets. Researchers have conducted user studies to understand users' conceptual models of spreadsheets to identify how the cognitive process can affect error rates \cite{panko2016we}, to see how users navigate large spreadsheets \cite{mittermeir2002finding}, to evaluate how multiple users interact with a single spreadsheet \cite{nardi1990ethnographic}, and to characterize the strengths and weaknesses of spreadsheets \cite{nardi1990spreadsheet, hendry1994creating}. Other studies have focused on errors: Powell et al. explore different types of errors that occur and how they can be minimized \cite{powell2008critical}, while others study real spreadsheets to discover errors \cite{aurigemma2014evaluating, kruck2001spreadsheet}. Our approach is instead to identify scalability problems in spreadsheets by exploring troubleshooting posts on an online forum.

\subsection{Using online communities as a data source}
The availability of diverse and large amounts of  online community data has led researchers to mine this data to answer research questions \cite{jang2011youtube, anthony2013analyzing, kulshrestha2017quantifying, kim2016power, keelan2007youtube}. While this method may bias the user sample to users with internet access and a level of technology savviness,  prior works have successfully created rich characterizations of users via this approach. In addition to the papers mentioned in the introduction~\cite{anthony2013analyzing, kim2016power, KeertipatiSL16}, Kulshrestha et al. \cite{kulshrestha2017quantifying} measured the political bias of an individual Twitter search result by extracting features from the Twitter user's account, while Keelan et al. \cite{keelan2007youtube} extracted Youtube videos to measure the sentiment (positive/negative) surrounding immunization. 

%% file: 01.5-reddit.tex
\section{Characterizing Reddit}

Reddit is a website that hosts a variety of forums called subreddits. Each subreddit is a forum dedicated to a specific topic and is named /r/topic\_name. Within these forums, users can post questions or notes and can comment on each others' posts, leading to discussions. Reddit has an API \cite{redditapi} to access data from the site, making it ideal to use for those interested in automating the scraping and categorizing of data.

\subsection{The Forum}
The /r/excel forum has been around for over 8 years, and as of September 2017, had over 70,000 subscribers \cite{excelmetrics}. Note that a subscriber is someone who subscribes to (follows) the forum; anyone with a Reddit account can read, post, or comment in the /r/excel forum without following it. Reddit recently removed its statistics regarding the traffic of subreddits to improve user privacy \cite{redditchanges}, resulting in the number of subscribers being the most valuable statistic publicly available. Note also that this forum is monitored (meaning a moderator can take down posts as they see fit), and one of the moderators of the forum disclosed in a post entitled "Please welcome our new Corporate Overlords" that the subreddit is involved in the Excel Influencer Program \cite{welcome}.

\subsection{The Users of the Forum}
Reddit does not collect any information from the user other than a username, password, and email, so it is difficult to characterize the type of users that frequent Reddit, never mind those that visit the /r/excel subreddit. However, in their posts and comments, some users share their experience level with Excel. Often users who ask questions begin with a statement of their unfamiliarity with Excel. Others state that they use Excel frequently for work but that they need help performing a new and/or complex operation. The users who answer questions typically do not state their credentials, but often their complex solutions indicate a high level of experience with Excel.

In some cases, there are tools other than Excel that are useful for managing (particularly large amounts of) data like Microsoft's Access software \cite{access} or relational databases. Users of the forum (both post makers and commenters) showed varying degrees of knowledge about these alternatives. In some cases, the user wanted to know if a tool was more appropriate than Excel. In others, the users specifically said they knew other programs would perform better, but they were forced to use Excel. Sometimes the user did not mention Access or databases, but was recommended by commenters to use a database as opposed to a spreadsheet. Out of the 712 posts we collected, 89 mentioned one or more of the terms \textit{``Access'', ``SQL'', } or \textit{``database''} in the post body or comments.

\subsection{The Uses of Excel}
In the 712 posts, the uses of Excel seemed to fall into two overarching categories: Excel for personal use, and Excel for professional use. Traditionally, in both of these areas, we think of spreadsheets as being used for record keeping of data like addresses and emails, time trackers and schedules, or financial information and budgets. There were many posts regarding these topics in both personal and professional settings. However, the more unusual uses of Excel were impressive, detailed next.

Regarding personal uses, several users asked about how to keep track of and calculate \emph{sports statistics}; fantasy football was particularly popular. One user wanted to design a spreadsheet that automatically organizes a table tennis tournament. Another wanted to create an ELO rating system for professional tennis players. The games mentioned also included virtual ones. Several people asked questions about keeping track of \emph{statistics for video games} including Pokemon and League of Legends. Some users went even further and \emph{created fully functional versions of popular games} like Tic-Tac-Toe or Blackjack in Excel. Another popular personal use was \emph{keeping track of health information} like food consumption and body measurements in Excel.

Professionals from a variety of industries used Excel spreadsheets for work. Unsurprisingly, many users' questions related to time tracking and inventory or office management. Several posts dealt with \emph{financial information}, including two people who were pulling live stock statistics from the web. Manufacturing industry professionals created spreadsheets that serve as \emph{specialized calculators}. One user  created a sheet that accepts input from a bar code scanner and organizes data accordingly. \emph{Real estate industry} professionals used Excel to track renovation progress at tens of locations or to modeling rental income. Several types of \emph{scientists utilize Excel's data analytics features}. Two biologists asked questions regarding spreadsheets they used to organize animal tracking data. Other scientists used Excel to create models of hydraulics and analyze a set of lunar data. The medical industry uses Excel not just to \emph{keep track of things like patient diagnoses and diets}, but also for \emph{medical research}. One user in particular created graphs in Excel that analyzed a set of brain wave data.

Overall, we found posts that touched on the following users/industries: real estate, finance, manufacturing, accountants, biologists, doctors and medical staff, teachers, researching students and professors, family management, and quantified self. Note that this is information from users who identified the context of their work. The majority of posts simply listed their problem without giving context as to where the spreadsheet is used.

%% file: 02-relatedwork.tex
\section{Methodology}
One goal of this work is to understand the breadth of issues users experience while using Excel, specifically issues related to scalability. To accomplish this, we chose to analyze posts from the Excel forum ``/r/excel'' on Reddit. There are other spreadsheet related forums such as /r/spreadsheets and /r/sheets. However, as Microsoft Excel is one of the largest players in the spreadsheet industry, we limited our scope to the Excel forum.

\subsection{Constructing and coding a Dataset from Reddit}

We collected two datasets of posts from the /r/excel forum using the Reddit API: one dataset to understand the general scope of Excel challenges and a second dataset to understand scalability issues specifically. 

\subsubsection{The random dataset}
{\bf Collection.} To collect our first dataset, we used the Reddit API's ``random'' function, fetching 50-100 posts with each call and collected 278 posts in May of 2017 over the period of two weeks. With this data we aimed to understand the breadth of the issues users encountered. 

{\bf Coding.}  We used open coding to code our random posts. Two authors individually examined 50-100 fetched posts at a time and developed criteria to separate the posts into themes (and then sub-themes). The two then compared their themes to create a codebook to create a common set of themes. They repeated the process of fetching and  categorizing posts until no new themes emerged. 

Most of these posts related to a user's uncertainty of how to perform operations (e.g., how can one convert columns into rows) or described a behavior not meeting expectations (e.g., "AND" operation did not join two clauses in an array formula correctly, resulting in no data returned). Two posts described scalability issues, where the dataset size caused undesired behavior: in one, a formula calculated for 30 minutes before crashing, and in the second, Excel froze every time the screen shifted.
We were intrigued by the scalability issues because (1) to date, they have not been chronicled in existing work, and (2) the outcomes of anecdotal cases did not have solutions. We, therefore collected a second dataset to explore scalability (see below).

{\bf Establishing Reliability.} For both this and the following set, after creating the categories, two of the authors coded sets of 10\% of the posts from each dataset separately until they achieved a Cohen's kappa of at least .78. They then proceeded to code the entirety of the dataset, comparing 5\% of all post codes every 85 posts (20\% of all posts), ensuring that they maintained a Cohen's kappa of at least .7.

\subsubsection{The search term dataset}
{\bf Collection.} We used the Reddit API and searched for scalability posts via keywords within the same /r/excel forum. Adapting the methodology used in \cite{koutra2015events, kulshrestha2017quantifying, kim2016power}, we finalized the set of search terms in the following manner: (1) From the randomly collected posts, we identified scalability posts and their keywords. (2) Using the keywords found in (1), we searched for additional scalability posts. We then extracted new scalability keywords from these posts. We repeated this process until no new keywords/phrases were identified. The final list of scalability keywords was \textit{``big'', ``crash'', ``forever'', ``freeze'', ``lag'', ``large'', ``long time'',} and \textit{``slow''}.

Using our final set of eight keywords/phrases, we created a dataset of 434 posts by collecting the first 100 posts returned by each keyword or until the API call ceased to return results, whichever happened first, from July through August of 2017. Typically, approximately 70 posts were returned, but some keywords returned as few as nine. 

While this approach was designed to filter scalability issues from the forum as a whole, we found that people also addressed Excel ``crashing'' or ``freezing'' in other situations (e.g., ``freeze panes" or ``crash course in Excel").  To capture posts that addressed the scalability issues, we created a codebook to separate scalability issues from non-scalability issues.  We stopped collecting data when no new themes emerged in our coding of problem area themes.

{\bf Coding for Theme.} We adapted the coding 
process we used for the random dataset to code this search term dataset. We ran each search term separately and coded resulting posts in batches as they arrived.  Two of the authors developed criteria to separate the posts into themes (and then sub-themes). The two then compared their themes to create a codebook to create a common set of themes. They repeated the process of fetching and  categorizing posts until no new themes emerged. While we started with a blank slate after coding the random dataset, we  found that the posts essentially fell into the same themes as the randomly fetched posts. The posts that did not fall into the four main themes were placed in a \textit{miscellaneous} category.

{\bf Coding for Scalability.} As described earlier, we separated the scalability related posts from the non-scalability related posts. The authors created the following criteria for this coding: a scalability issue is defined as \textit{poor} behavior within an Excel workbook that contains a \textit{large amount of data}. Specifically, poor behavior is characterized as Excel lagging, not responding, crashing, or taking long enough to respond that the user declared the spreadsheet unusable. The inclusion criteria for scalability posts were: (1) the user specified observing thousands of rows of data, (2) the user's file size was 5MB \footnote{This is the size where users began experiencing ``lag'' and ``freezing'' issues.} or larger, (3) the user did not quantitatively specify how large their data was, but they expressed suspicion that the issues were due to the dataset size, (4) a user could successfully import a small amount of data, but failed when the data size increased.
 
 We coded the same dataset thematically and for scale. For both codings of the search term dataset,  we established coding reliability in the same manner as we described in the random dataset coding section.
 From the 434 search term posts, 81 posts addressed scalability concerns, bringing the combined number of posts dealing with scalability issues to 83 out of 712.

%% file: 03-methodology.tex
\section{The Themes}

\begin{figure*}
    \centering
    \includegraphics[scale=0.50]{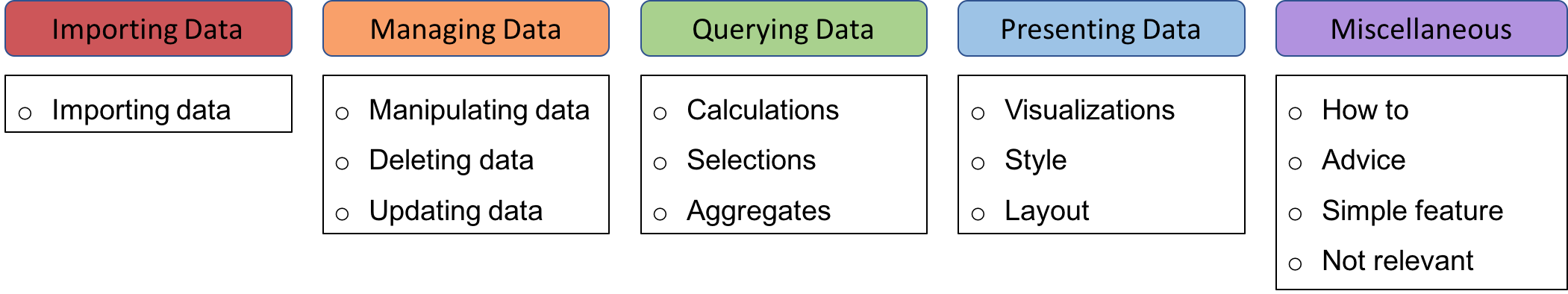}
    \caption{A hierarchical representation of the themes and their sub-themes.}
    \label{fig:themes}
\end{figure*}

We made two observations which laid out the architecture for our thematic coding. First, the posts were overwhelmingly questions, as opposed to statuses or updates. Second, the majority of the questions dealt with {\em four themes: importing data, managing data, querying data, and presenting data.} While the majority of posts fell into these four themes, the remainder were placed in a Miscellaneous theme. These posts generally asked broader, questions than other posts, as we discuss further below.

Our four main themes were further categorized into sub-themes. We introduce the themes and sub-themes below with a brief description of the posts they contain, followed by a table presenting paraphrased quotes (to preserve anonymity) of posts from a selection of sub-themes. Each post received one of the following labels as well as a yes or no label indicating whether it related to scalability to better understand both the post issue whether the issue was affected by scalability. The pair of labels allowed us to see the operation areas affected by scalability. To illustrate how the ``scalability'' code changes the nature of the problems users encountered, we present several posts in Table~\ref{fig:post_excerpts} with their scalability code.

\subsection{Importing Data}
This theme includes posts related to importing data from sources such as CSV files, PDFs, websites, and other Excel workbooks. 

\subsection{Managing Data}
This theme emerged from posts that discussed cleaning and updating data after it has been imported. This theme included the three sub-themes below.

{\bf Manipulating Data.} This sub-theme includes questions about actions required to transform data into a analyzable state. (e.g., cleaning data after import and changing the format of data for a non-aesthetic purpose--changing the format of a date so it is consistent with existing data and formulas in the sheet.)

{\bf Deleting Data.}
This sub-theme includes questions that ask how to delete or hide data from view.

{\bf Updating Data.}
This sub-theme includes questions asking how to update existing entries in a workbook or how to add more data to a workbook, e.g., adding a fixed value to all of the entries in a column or adding new entries to a list of records in a sheet.

\subsection{Querying Data}
This theme includes posts related to extracting information from the data in a workbook.

{\bf Calculations.}
This sub-theme includes questions on how to create new values from existing data in the spreadsheet. Note that simply selecting existing data from the sheet does \textit{not} qualify as a calculation; \textit{new} data is created.

{\bf Selections.}
This sub-theme includes posts about finding and/or selecting pieces of information from a workbook (e.g., questions whose main topic is VLOOKUP). Other common examples discuss operations such as finding duplicates or finding which records fall between two dates.

{\bf Aggregations.}
This sub-theme includes posts that ask questions about grouping all records of a sheet into themes, counting entities (often via COUNTIF), summing entities, or using pivot tables. The main distinction between \textit{Aggregations} and \textit{Selections} is that \textit{Selections} deal with selecting data that meets certain conditions while disregarding the remainder of the data, whereas \textit{Aggregations} deal with organizing/categorizing \textit{all} of the data in a sheet. A  \textit{selection} functions as a drop down filter. An \textit{Aggregation} functions similarly to a SQL groupby statement, as it organizes all of the data in a table.

\subsection{Presenting Data}
This theme includes posts relating to how the data is presented to the Excel user (and not necessarily for the purpose of creating a formal presentation).

{\bf Visualizations.}
This sub-theme includes posts that relate to the creation of charts, graphs, and tables. The questions address creating templates, changing the style of a chart, or selecting the correct data to be used in a chart.

{\bf Style.}
This sub-theme includes posts that deal with the style of the Excel sheet or the style of the data within the sheet. Examples of style actions in the sheet include  highlighting, outlining cells, and changing row/column widths. Actions that deal with the style of the data include changing font properties or the number of decimal places displayed. \techreport{One important distinction with the style of the data is that it is done for an aesthetic purpose; it does not change the traits of the data that would influence queries.}

{\bf Layout.}
This sub-theme includes posts that deal with the position or layout of data within the workbook. Questions asking how to move data \textit{within the same workbook} also fall into this theme. Other common questions  include moving all data in a certain manner (e.g., up one row or converting rows to columns).

\subsection{Miscellaneous}
These posts were significant enough to warrant coding, but did not fall into one of the aforementioned themes.

{\bf How-To or Isn't Working.}
This sub-theme included posts that either ask how to do a task (not present in the previous four themes) or ask why a simple feature in Excel does not meet user expectations.
Often these formats of questions are combined into one---"X is not working, how would I solve this problem". Creative uses of Excel (e.g., creating games within a spreadsheet) fell in this sub-theme.

{\bf Advice.}
This sub-theme includes posts that ask for general advice about Excel functionality. If a post was specifically asking about aggregating data, for example, we included it in the \textit{aggregate} theme. However this theme captured questions that were broad in scope---for example, ``how do I structure a large amount of data in a spreadsheet''.

{\bf Simple Expected Features.}
This sub-theme includes posts that contain descriptions of simple features that did not function as expected. We define simple features as those  required for Excel to properly function as data storage mechanism (e.g., opening/closing the application, adding a row to a sheet, saving the data, and scrolling).  

{\bf Not Relevant.}
This sub-theme includes posts that were not relevant to our case study addressing spreadsheet needs and desires. These posts included advertisements or requests for software alternatives to Excel.

\begin{table}[h]
    \centering
    \includegraphics[scale=0.45]{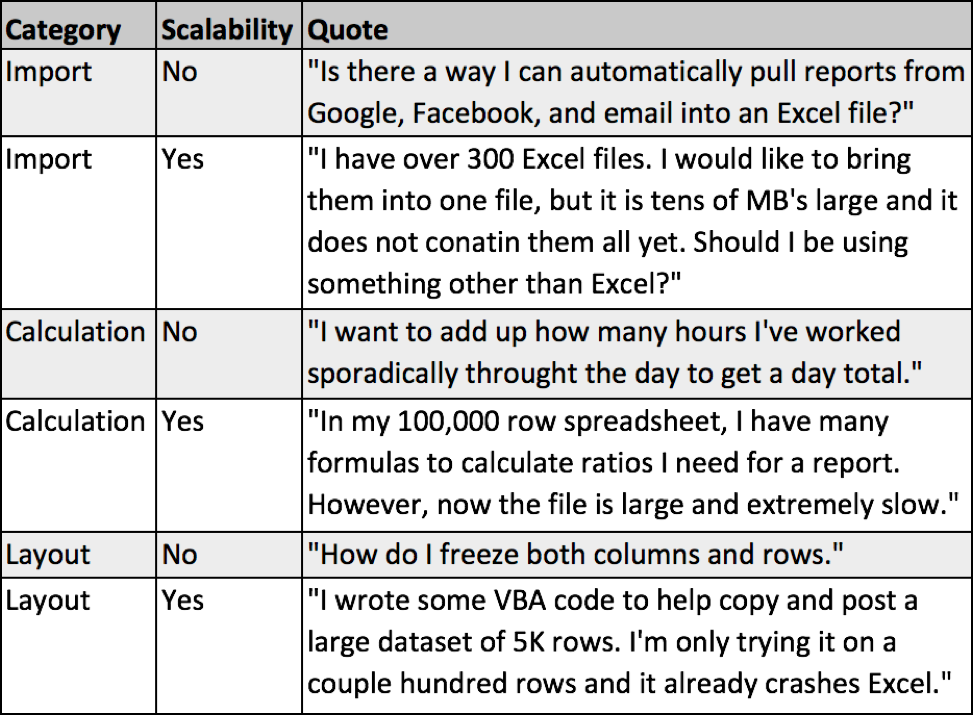}
    \caption{Excerpts from collected posts.}
    \label{fig:post_excerpts}
\end{table}

%% file: 04-findings.tex
\section{Results and Discussion}

\begin{figure}[hbt]
    \centering
    \includegraphics[scale=0.50]{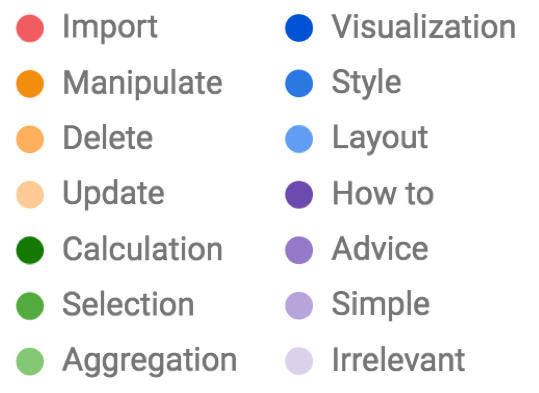}
    \caption{The legend for the following charts.}
      \label{fig:legend}
\end{figure}

\begin{figure}[hbt]
    \centering
    \includegraphics[scale=0.30]{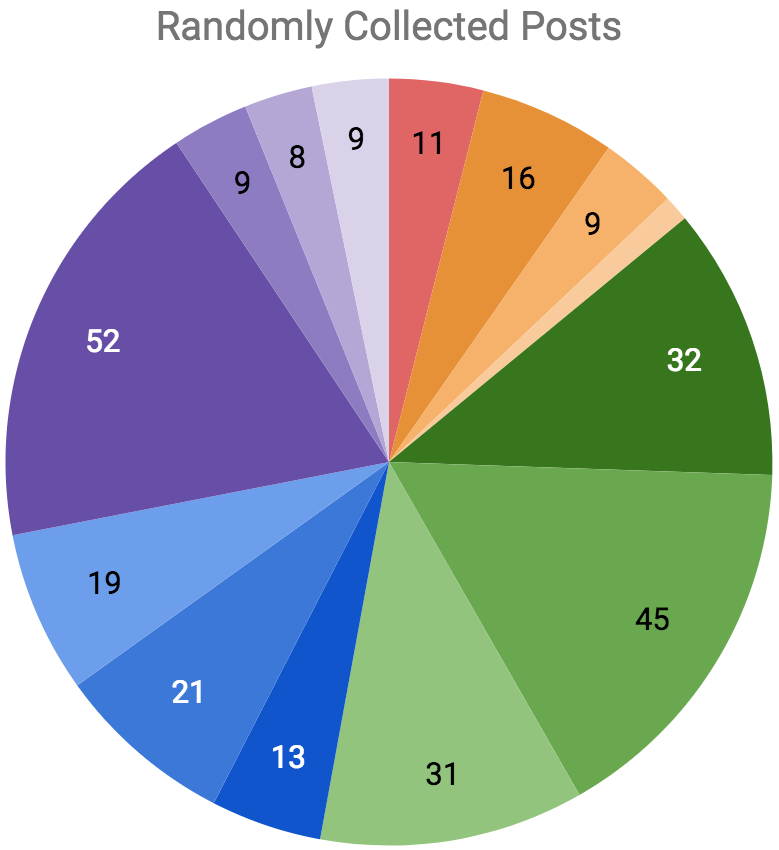}
    \caption{The distribution of the random posts (legend in figure Figure~\ref{fig:legend}).}
      \label{fig:random_post_distributions}
\end{figure}

\begin{figure}[hbt]
    \centering
    \includegraphics[scale=0.30]{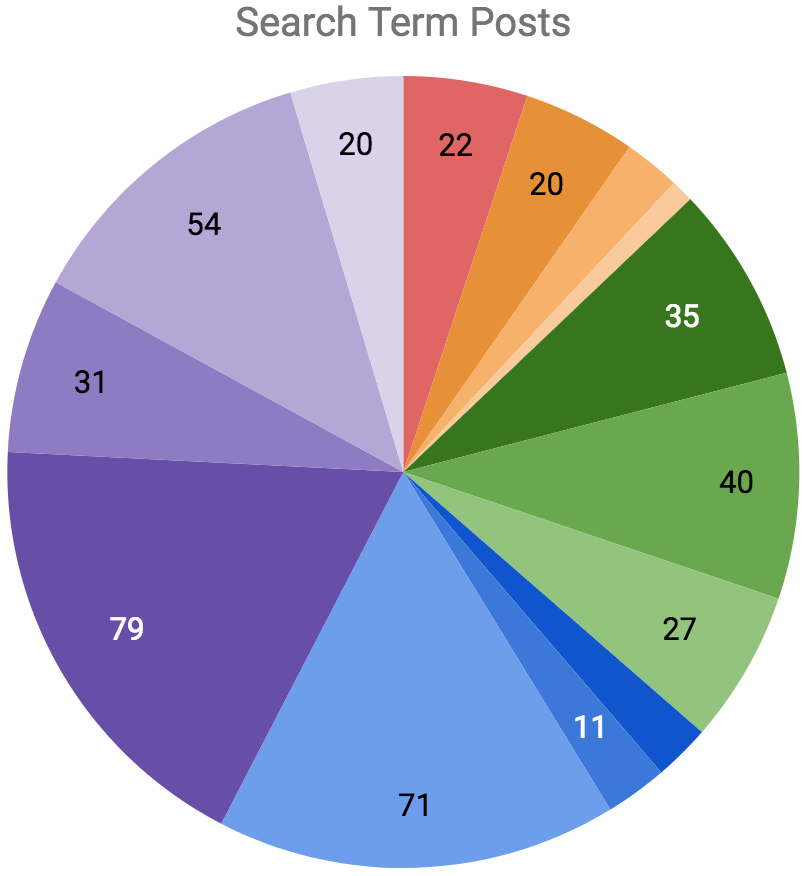}
    \caption{The distribution of the search term posts (legend in  Figure~\ref{fig:legend}).}
      \label{fig:search_term_distributions}
\end{figure}

\begin{figure}[hbt]
    \centering
    \includegraphics[scale=0.33]{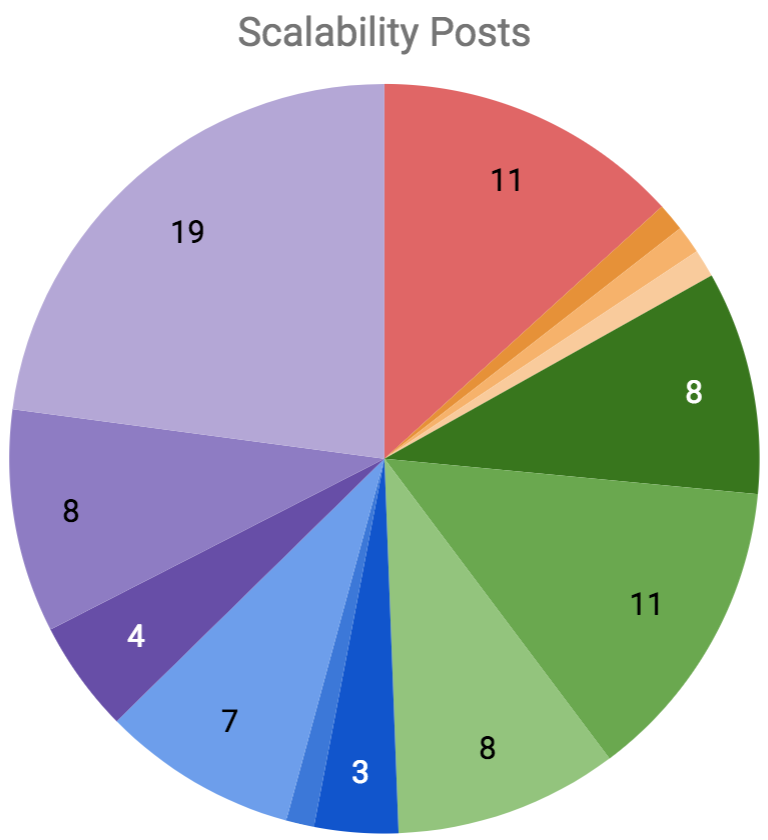}
    \caption{The distribution of the scalability posts (legend in Figure~\ref{fig:legend}).}
      \label{fig:scalability_distribution}
\end{figure}

The main goal of this work is to present the landscape of scalability issues present in Excel as captured in a popular subreddit forum. We first present characteristics of the posts before describing distribution of results in categories.

 Figure~\ref{fig:random_post_distributions} presents the distribution of the set of 278 random posts, Figure~\ref{fig:search_term_distributions} presents the distribution of the 434 search term posts, and Figure~\ref{fig:scalability_distribution} shows the distribution of the posts we categorized as relating to scalability. 83 of the 712 posts related to scalability. 81 came from the search terms posts, while the other two came from the random posts. We now discuss the characteristics of the posts and the conclusions we made.

\subsection{Results Regarding All Posts}
We collected 712 posts in total. Of these posts, 644 posts had unique authors. 21 posts had no author \footnote{We have not found a source for why a post can have no author. We assume that this occurs when a user makes a post and then deletes their account.}. We were interested to see if authors of comments were similarly diverse, or if there were a few users that commented on many posts. We found that there were 315 unique comment authors-- one of which was "no author". For these 712 posts, the most unique posts that one user commented on was 8. The most comments that one user made was 18 across a total of 7 different posts. Again, there does not appear to be a small number of users that dominate the comments.

The average length of a post (not including the comments) was 135 words, and the average number of comments was 7-- note that this includes all comments posted by users and bots. However, 42 posts received no comments. On this subreddit, users can specify that their post is a question which starts out in the "unresolved state". The post maker can then wait for other users to comment and suggest solutions. If a solution works, they comment "solution verified" and the post is marked as resolved. If the user fails to mark the post as resolved within a certain amount of time, a bot comments reminding them to mark a solution as verified. 91 posts had the "solution verified" comment.

Reddit allows users to "upvote" and "downvote" posts. On the /r/excel subreddit, when a user tries to downvote a post, a box pops up indicating that the user should comment why they are downvoting and says specifically "this is NOT a disagree button". This comment likely indicates that the forum encourages the number of upvotes and downvotes to reflect the quality of the post as opposed to if they like/agree with it. The average upvote to downvotes ratio of the 712 posts was 84.2\%, indicating that the majority of posts users believe to be of high quality.

Traditional methods of determining quality of posts and discussions include looking at the thread depth of posts ~\cite{whittaker2003dynamics}. Here, we measure thread depth as the height of the comment tree, where the original post has thread depth 0, a direct reply to the original post has depth 1, and so on. The average thread depth of the posts in the forum was 3, with the maximum and minimum thread depths being 10 and 1. One of the posts with a thread depth of 1 stated that Excel crashed every time it was closed and listed the solutions that had already been tried. A post that had a thread depth of 10 asked a more general question. For example, one post that had a thread depth of 10 asked how to best encrypt a spreadsheet.

\subsection{Results Specific to Scalability Posts}

As we specifically wanted to study the circumstances that lead to scalability issues, we analyzed the post content and comments of scalability posts in depth.

\subsubsection{File Size}
We were curious as to what sized file are susceptible to scalability issues. Microsoft specifies data limits in the Excel documentation \cite{limits}. Specifically, Excel can handle no more than 1,048,576 rows by 16,384 columns. We expected that worksheets with over 1 million rows would start to have scalability issues. Indeed, the largest spreadsheet we saw in our scalability posts contained about 5 million rows, and Excel was slow to partially open it. The smallest sheet that we classified as having scalability issues had 18 thousand rows. This user wanted to perform multiple conditional if statements on the data, but needed a more time efficient solution. As the solution of using conditional if statements would have worked on a smaller dataset, we classified it as a scalability issue.

\subsubsection{Suggested Solutions}
Every scalability post included a question and most received more than one suggested solution from other Reddit users. We analyzed the solutions that were frequently recommended in each of the importing data, managing data, querying data, and presenting data themes. The following solutions were common across all of these four themes:
\begin{enumerate}
    \item \textbf{Use a database.} This common solution was suggested by multiple people in 38 of the 83 scalability posts. Reddit users commented that, once a large enough amount of data is stored in the spreadsheet, a database is the best way to ensure that one can still manage and query the data in a reasonable amount of time. Though commenters indicated using a database is the best solution, they offered other suggestions, if using Excel is unavoidable. For example, PowerQuery and PowerPivot, Excel extensions that are meant to process large amounts of data more easily, are not as powerful as a database, but are still better than using basic Excel. Otherwise, users are recommended to perform tedious operations like breaking up the data into chunks and processing the smaller chunks one at a time so as not to overwhelm the system.
    
    \item \textbf{Turn of automatic calculations.}
    This was a suggested solution for 12 of the 83 scalability posts. Excel is set by default so that once a user updates a cell \emph{c}, the system will recalculate any cells that depend on \emph{c} immediately. This can be very time consuming, especially when an update to one cell triggers a chain of recalculations. Turning off automatic calculations results in Excel waiting to recalculate formulas after an update until the user manually selects to recalculate. This does not necessarily make the calculations faster, but will allow the spreadsheet to remain interactive while changes are made.
    
    \item \textbf{Save Excel files as .XLSB files.} This was a suggested solution for 12 of the 83 scalability posts. By default, Excel files are saved as .XLS or .XLSX files. If instead a spreadsheet is saved as a .XLSB file, it is stored in binary format. The effects of the binary format include a smaller file size and shorter loading and saving times.    
    
    \item \textbf{Use as little conditional formatting as possible.} This was a suggested solution for 10 of the 83 scalability posts. Conditional formatting is the adjusting the formatting of a cell based on a conditional statement. For instance, highlight each cell in column C if its value is less than 50. Multiple users report that conditional formatting slows down an Excel file, particularly if it affects many rows.
    
    \item \textbf{Avoid using volatile functions.} This was a suggested solution for 10 of the 83 scalability posts. Functions like OFFSET are considered volatile, meaning that any update to any cell in the spreadsheet triggers a recalculation of this formula. Obviously, this can become very expensive unnecessarily.
        
    \item \textbf{Avoid using computationally intensive functions.} This was a suggested solution for 9 of the 83 scalability posts. Computationally intensive functions include VLOOKUP and COUNTIF. Users indicate that these functions, though sometimes necessary, oftentimes slow a spreadsheet down, particularly if the sheet has a lot of data. One alternative to VLOOKUP is INDEX MATCH. INDEX MATCH is preferred to VLOOKUP as INDEX MATCH often examines fewer cells in total than VLOOKUP. 

\end{enumerate}

 We drew two main conclusions:

\textit{\textbf{1. Users understand the capabilities of Excel, but not how to operationalize them.}}
As a whole, the results stress the importance of Excel as a data storage tool and as a data processing engine (these posts account for 63.2\% of all posts collected). Note the large difference between the number of posts that fell in the \textit{querying data} (29.5\% of posts) theme compared to \textit{importing} and \textit{manipulating data} combined (13.3\% of posts). This indicates that users are generally comfortable with using Excel as a tool for storage, but they struggle when it comes to applying functions which allow them to extract useful information from their data.

In the majority of posts, users state what they would like Excel to do, and ask how to obtain this result. People do not underestimate what can be accomplished with Excel. The questions span the whole gamut of data analysis, from data collection (crawling the web with VBA code) to consumption of information (visualizations) and everything in between. Many of the posts in the \textit{How to} category deal with people trying to accomplish creative tasks with Excel. One user was trying to implement Tic-Tac-Toe. Another was trying complete their RSA encryption homework in a spreadsheet.

These results together indicate that we need to make these complex and powerful capabilities more accessible to users. The technical knowledge of post makers ranged from people who were not familiar with VBA and wanted to avoid a solution using VBA to those who used it frequently. \techreport{To cater to both types of users, perhaps more research can be given into spreadsheet tutorials and documentation.}

Many of the posts tell a story of an often complex task the user wants to complete. Context is given, a goal is explained, and ideas of implementation are shared. Often these posts reflect an understanding of how to perform operations in Excel, but express confusion with stringing multiple operations together to perform a complete task. While documentation is very useful for understanding how a specific function works, it is not much help in understanding how to build a solution to a large problem out of basic functions. More research in the area of helping users understand how to integrate functions together would be of great use to many of the post makers on the /r/excel subreddit.

\textit{\textbf{2. Scalability issues affect a wide range of operations in spreadsheets.}}
Every theme and sub-theme contains posts relating to scalability. Analyzing this distribution can help prioritize which operations are the most important to adapt to handle large quantities of data. For example, a large amount of the scalability issues deal with either importing data or querying data. These two areas in particular require further attention.

Notice that scalability issues cause a large number of the issues in the \textit{simple expected feature} posts. This indicates that as data sets become larger, even the simplest of features in Excel cease to respond in a timely manner. Without these features, as the post makers comment, Excel is essentially useless. Users turn to other options like databases, but some are reluctant to use them because they are not familiar with SQL. Ideally, these users who are not familiar with databases should be able to use the spreadsheet interface they are comfortable with, while utilizing the processing power and scalability of a database.

Moreover, from an expressiveness point of view, many of the users' issues could be solved by a simple SQL statement. Multiple times, we witnessed a user trying to create a complex formula to accomplish what could be done with a simple select statement. For example, one post maker asked "I want to be able to input two dates, and have shown the text and date for every date between those. I think i[sic] need to use some sort of array formula, but I'm not sure how (the exact formula, not array as a whole)". The suggested solution posted was to use the following formula:

{\tt =IFERROR(INDEX(B:B,SMALL(IF((\$B\$5:\$B\$80>=\$C\$34) * (\$B\$5:\$B\$80<=\$F\$28), ROW(\$B\$5:\$B\$80)), ROWS(\$B\$5:\$B5))), "")}

Not only is this formula hard to parse and understand, but also the post maker 
was not able to conceive of this formula themselves. They further could not alter the suggested formula for their task and needed assistance to organize the results. The solution to this problem is much simpler and cleaner in SQL:

{\tt SELECT text, date FROM table WHERE date > date1 AND date < date2}

Multiple users similarly try to create a complex formula using VLOOKUP in order to complete what at its basic form is a join.
Similar observations could be made of many of the questions in the {\em querying data} theme---that they could be expressible as simple SQL queries (with joins, group-by, and selections)~\cite{bakke2011spreadsheet,nandi2013interactive,abouzied2012dataplay}. Using SQL queries would help users avoid using computationally intensive formulae (as was suggested by Reddit users) like COUNTIF, SUMIF, and VLOOKUP. 

%% file: 05-evaluation.tex
\section{Limitations}
Our methodology poses some limitations. Our collection approach using the ``random'' API call is described as a ``the serendipity button,''  indicating that it returns a random post from the forum. However, verifying that it is truly random is challenging.  In our results, we see a bias towards more recent posts. We also have a self-selection bias in the posters that contribute to the subreddit we studied.

%% file: 06-discussions.tex
\section{Conclusions}
In our case study, we set out to apply a method of utilizing online forum posts to understand usability issues people face with spreadsheets. 
The majority of posts in the subreddit were Excel users asking questions about the four main steps needed to process data: importing, managing, querying, and presenting---we found that all four are important. Furthermore, not only are scalability issues very real problems plaguing those who want to analyze large amounts of data, but these issues touched on all four areas of actions users take to process data. This motivates the need for developing a spreadsheet application which continues to perform as expected on very large volumes of data~\cite{bendre2017towards}.

%% file: 07-conclusion.tex
\begin{table}
\begin{center}
 \begin{tabular}{||c c c||} 
 \hline
 Name & Num Posts & Percentage \\ [0.5ex] 
 \hline\hline
     Import & 11 & 4.0 \\
     \hline
     Manipulate  & 16  & 5.8  \\
     \hline
     Delete  & 9  & 3.2  \\
     \hline
     Update  & 3  & 1.1 \\
     \hline
     Calculation  & 32  & 11.5 \\
     \hline
     Selection  & 45  & 16.2 \\
     \hline
     Aggregation  & 31  & 11.2 \\
     \hline
     Visualization  & 13  & 4.7 \\
     \hline
     Style  & 21  & 7.6 \\
     \hline
     Layout  & 19  & 6.8 \\
     \hline
     How to  & 52 & 18.7 \\
     \hline
     Advice  & 9  & 3.2 \\
     \hline
     Simple  & 8  & 2.9 \\
     \hline
     Irrelevant  & 9  & 3.2 \\ [1ex] 
 \hline
\end{tabular}
\end{center}
 \caption{Accessible representation of distribution of random posts (figure 3).}
\end{table}

\begin{table}
\begin{center}
 \begin{tabular}{||c c c||} 
 \hline
 Name & Num Posts & Percentage \\ [0.5ex] 
 \hline\hline
     Import & 2 & 5.1 \\
     \hline
     Manipulate  & 20  & 4.6  \\
     \hline
     Delete  & 10  & 2.3  \\
     \hline
     Update  & 4  & .9 \\
     \hline
     Calculation  & 35  & 8.1 \\
     \hline
     Selection  & 40  & 9.2 \\
     \hline
     Aggregation  & 27  & 6.2 \\
     \hline
     Visualization  & 10  & 2.3 \\
     \hline
     Style  & 11  & 2.5 \\
     \hline
     Layout  & 71  & 16.4 \\
     \hline
     How to  & 79 & 18.2 \\
     \hline
     Advice  & 31  & 7.1 \\
     \hline
     Simple  & 54  & 12.4 \\
     \hline
     Irrelevant  & 20  & 4.6 \\ [1ex] 
 \hline
\end{tabular}
\end{center}
 \caption{Accessible representation of distribution of search term posts (figure 4).}
\end{table}

\begin{table}
\begin{center}
 \begin{tabular}{||c c c||} 
 \hline
 Name & Num Posts & Percentage \\ [0.5ex] 
 \hline\hline
     Import & 11 & 13.3 \\
     \hline
     Manipulate  & 1  & 1.2  \\
     \hline
     Delete  & 1  & 1.2  \\
     \hline
     Update  & 1  & 1.2 \\
     \hline
     Calculation  & 8  & 9.6 \\
     \hline
     Selection  & 11  & 13.3 \\
     \hline
     Aggregation  & 8  & 9.6 \\
     \hline
     Visualization  & 3  & 3.6 \\
     \hline
     Style  & 1  & 1.2 \\
     \hline
     Layout  & 7  & 8.4 \\
     \hline
     How to  & 4 & 18.7 \\
     \hline
     Advice  & 8  & 9.6 \\
     \hline
     Simple  & 19  & 22.9 \\ [1ex] 
 \hline
\end{tabular}
\end{center}
 \caption{Accessible representation of distribution of scalability posts (figure 5).}
\end{table}

%% file: main.bbl

\begin{thebibliography}{00}


\ifx \showCODEN    \undefined \def \showCODEN     #1{\unskip}     \fi
\ifx \showDOI      \undefined \def \showDOI       #1{{\tt DOI:}\penalty0{#1}\ }
  \fi
\ifx \showISBNx    \undefined \def \showISBNx     #1{\unskip}     \fi
\ifx \showISBNxiii \undefined \def \showISBNxiii  #1{\unskip}     \fi
\ifx \showISSN     \undefined \def \showISSN      #1{\unskip}     \fi
\ifx \showLCCN     \undefined \def \showLCCN      #1{\unskip}     \fi
\ifx \shownote     \undefined \def \shownote      #1{#1}          \fi
\ifx \showarticletitle \undefined \def \showarticletitle #1{#1}   \fi
\ifx \showURL      \undefined \def \showURL       #1{#1}          \fi

\bibitem{limits}
 2017.
\newblock Excel specifications and limits.
\newblock \url{goo.gl/PiWptB}.   (2017).
\newblock
\newblock
\shownote{Accessed: September 24, 2017.}


\bibitem{welcome}
 2017.
\newblock Please welcome our new Corporate Overlords.
\newblock
  \url{https://www.reddit.com/r/excel/comments/3lusom/please_welcome_our_new_corporate_overlords/}.
    (2017).
\newblock
\newblock
\shownote{Accessed: September 24, 2017.}


\bibitem{redditapi}
 2017a.
\newblock Reddit API Documentation.
\newblock \url{https://www.reddit.com/dev/api/}.   (2017).
\newblock
\newblock
\shownote{Accessed: September 15, 2017.}


\bibitem{excelmetrics}
 2017.
\newblock /r/excel metrics (Microsoft Excel | Help and Support with your
  Formula, Macro, and VBA problems | A Reddit Community).
\newblock \url{http://redditmetrics.com/r/excel}.   (2017).
\newblock
\newblock
\shownote{Accessed: September 24, 2017.}


\bibitem{redditchanges}
 2017b.
\newblock Upcoming Changes: View counts, users here now and traffic pages.
\newblock
  \url{https://www.reddit.com/r/ModSupport/comments/6atvgi/upcoming_changes_view_counts_users_here_now_and/}.
    (2017).
\newblock
\newblock
\shownote{Accessed: September 24, 2017.}


\bibitem{abouzied2012dataplay}
{Azza Abouzied}, {Joseph Hellerstein}, {and} {Avi Silberschatz}. 2012.
\newblock \showarticletitle{DataPlay: interactive tweaking and example-driven
  correction of graphical database queries}. In {\em Proceedings of the 25th
  annual ACM symposium on User interface software and technology}. ACM,
  207--218.
\newblock


\bibitem{alexander2014microsoft}
{Michael Alexander}, {Jared Decker}, {and} {Bernard Wehbe}. 2014.
\newblock {\em Microsoft business intelligence tools for Excel analysts}.
\newblock John Wiley \& Sons.
\newblock


\bibitem{anthony2013analyzing}
{Lisa Anthony}, {YooJin Kim}, {and} {Leah Findlater}. 2013.
\newblock \showarticletitle{Analyzing user-generated youtube videos to
  understand touchscreen use by people with motor impairments}. In {\em
  Proceedings of the SIGCHI conference on human factors in computing systems}.
  ACM, 1223--1232.
\newblock


\bibitem{aurigemma2014evaluating}
{Salvatore Aurigemma} {and} {Ray Panko}. 2014.
\newblock \showarticletitle{Evaluating the Effectiveness of Static Analysis
  Programs Versus Manual Inspection in the Detection of Natural Spreadsheet
  Errors}.
\newblock {\em Journal of Organizational and End User Computing (JOEUC)\/}
  {26}, 1 (2014), 47--65.
\newblock
\showDOI{%
\url{http://dx.doi.org/10.4018/joeuc.2014010103}}


\bibitem{bakke2011spreadsheet}
{Eirik Bakke}, {David Karger}, {and} {Rob Miller}. 2011.
\newblock \showarticletitle{A spreadsheet-based user interface for managing
  plural relationships in structured data}. In {\em Proceedings of the SIGCHI
  conference on human factors in computing systems}. ACM, 2541--2550.
\newblock
\showDOI{%
\url{http://dx.doi.org/10.1145/1978942.1979313}}


\bibitem{bakke2016expressive}
{Eirik Bakke} {and} {David~R Karger}. 2016.
\newblock \showarticletitle{Expressive query construction through direct
  manipulation of nested relational results}. In {\em Proceedings of the 2016
  International Conference on Management of Data}. ACM, 1377--1392.
\newblock


\bibitem{bendre2017towards}
{Mangesh Bendre}, {Vipul Venkataraman}, {Xinyan Zhou}, {Kevin Chen-Chuan
  Chang}, {and} {Aditya Parameswaran}. 2017.
\newblock \showarticletitle{Towards a Holistic Integration of Spreadsheets with
  Databases: A Scalable Storage Engine for Presentational Data Management}.
\newblock {\em arXiv preprint arXiv:1708.06712\/} (2017).
\newblock


\bibitem{chambers2010struggling}
{Chris Chambers} {and} {Chris Scaffidi}. 2010.
\newblock \showarticletitle{Struggling to excel: A field study of challenges
  faced by spreadsheet users}. In {\em Visual Languages and Human-Centric
  Computing (VL/HCC), 2010 IEEE Symposium on}. IEEE, 187--194.
\newblock
\showDOI{%
\url{http://dx.doi.org/10.1109/VLHCC.2010.33}}


\bibitem{hendry1994creating}
{David~G Hendry} {and} {Thomas~RG Green}. 1994.
\newblock \showarticletitle{Creating, comprehending and explaining
  spreadsheets: a cognitive interpretation of what discretionary users think of
  the spreadsheet model}.
\newblock {\em International Journal of Human-Computer Studies\/} {40}, 6
  (1994), 1033--1065.
\newblock
\showDOI{%
\url{http://dx.doi.org/https://doi.org/10.1006/ijhc.1994.1047}}


\bibitem{jang2011youtube}
{Sun~Hee Jang}. 2011.
\newblock \showarticletitle{YouTube as an innovative resource for social
  science research}. In {\em Australian Association for Research in Education
  Conference (AARE 2011 Conference)}. Citeseer, 1--16.
\newblock


\bibitem{keelan2007youtube}
{Jennifer Keelan}, {Vera Pavri-Garcia}, {George Tomlinson}, {and} {Kumanan
  Wilson}. 2007.
\newblock \showarticletitle{YouTube as a source of information on immunization:
  a content analysis}.
\newblock {\em jama\/} {298}, 21 (2007), 2482--2484.
\newblock


\bibitem{KeertipatiSL16}
{Swetha Keertipati}, {Bastin Tony~Roy Savarimuthu}, {and} {Sherlock~A.
  Licorish}. 2016.
\newblock \showarticletitle{Approaches for prioritizing feature improvements
  extracted from app reviews.}. In {\em EASE}, {Sarah Beecham}, {Barbara
  Kitchenham}, {and} {Stephen~G. MacDonell} (Eds.). ACM, 33:1--33:6.
\newblock
\showISBNx{978-1-4503-3691-8}
\showURL{%
\url{http://dblp.uni-trier.de/db/conf/ease/ease2016.html##KeertipatiSL16}}


\bibitem{kim2016power}
{Jennifer~G Kim}, {Ha~Kyung Kong}, {Karrie Karahalios}, {Wai-Tat Fu}, {and}
  {Hwajung Hong}. 2016.
\newblock \showarticletitle{The power of collective endorsements: credibility
  factors in medical crowdfunding campaigns}. In {\em Proceedings of the 2016
  CHI Conference on Human Factors in Computing Systems}. ACM, 4538--4549.
\newblock


\bibitem{koutra2015events}
{Danai Koutra}, {Paul~N Bennett}, {and} {Eric Horvitz}. 2015.
\newblock \showarticletitle{Events and controversies: Influences of a shocking
  news event on information seeking}. In {\em Proceedings of the 24th
  International Conference on World Wide Web}. International World Wide Web
  Conferences Steering Committee, 614--624.
\newblock
\showDOI{%
\url{http://dx.doi.org/10.1145/2736277.2741099}}


\bibitem{kruck2001spreadsheet}
{Susan~E Kruck} {and} {Steven~D Sheetz}. 2001.
\newblock \showarticletitle{Spreadsheet accuracy theory}.
\newblock {\em Journal of Information Systems Education\/} {12}, 2 (2001),
  93--108.
\newblock
\showDOI{%
\url{http://dx.doi.org/10.4018/joeuc.2014010103}}


\bibitem{kulshrestha2017quantifying}
{Juhi Kulshrestha}, {Motahhare Eslami}, {Johnnatan Messias}, {Muhammad~Bilal
  Zafar}, {Saptarshi Ghosh}, {Krishna~P Gummadi}, {and} {Karrie Karahalios}.
  2017.
\newblock \showarticletitle{Quantifying search bias: Investigating sources of
  bias for political searches in social media}.
\newblock  (2017), 417--432.
\newblock
\showDOI{%
\url{http://dx.doi.org/10.1145/2998181.2998321}}


\bibitem{access}
{{Microsoft}}. 2017.
\newblock Access.
\newblock   (2017).
\newblock
\showURL{%
\url{https://products.office.com/en-us/access}}


\bibitem{mittermeir2002finding}
{Roland Mittermeir} {and} {Markus Clermont}. 2002.
\newblock \showarticletitle{Finding high-level structures in spreadsheet
  programs}. In {\em Reverse Engineering, 2002. Proceedings. Ninth Working
  Conference on}. IEEE, 221--232.
\newblock
\showDOI{%
\url{http://dx.doi.org/10.1109/WCRE.2002.1173080}}


\bibitem{nandi2013interactive}
{Arnab Nandi} {and} {Michael Mandel}. 2013.
\newblock \showarticletitle{The interactive join: recognizing gestures for
  database queries}. In {\em CHI'13 Extended Abstracts on Human Factors in
  Computing Systems}. ACM, 1203--1208.
\newblock


\bibitem{nardi1990ethnographic}
{Bonnie~A Nardi} {and} {James~R Miller}. 1990a.
\newblock \showarticletitle{An ethnographic study of distributed problem
  solving in spreadsheet development}. In {\em Proceedings of the 1990 ACM
  conference on Computer-supported cooperative work}. ACM, 197--208.
\newblock
\showDOI{%
\url{http://dx.doi.org/10.1145/99332.99355}}


\bibitem{nardi1990spreadsheet}
{Bonnie~A Nardi} {and} {James~R Miller}. 1990b.
\newblock {\em The spreadsheet interface: A basis for end user programming}.
\newblock Hewlett-Packard Laboratories.
\newblock


\bibitem{panko2016we}
{Ray Panko}. 2016.
\newblock \showarticletitle{What We Don't Know About Spreadsheet Errors Today:
  The Facts, Why We Don't Believe Them, and What We Need to Do}.
\newblock {\em arXiv preprint arXiv:1602.02601\/} (2016).
\newblock


\bibitem{powell2008critical}
{Stephen~G Powell}, {Kenneth~R Baker}, {and} {Barry Lawson}. 2008.
\newblock \showarticletitle{A critical review of the literature on spreadsheet
  errors}.
\newblock {\em Decision Support Systems\/} {46}, 1 (2008), 128--138.
\newblock
\showDOI{%
\url{http://dx.doi.org/10.1016/j.dss.2008.06.001}}


\bibitem{rajalingham2008classification}
{Kamalasen Rajalingham}, {David~R Chadwick}, {and} {Brian Knight}. 2008.
\newblock \showarticletitle{Classification of spreadsheet errors}.
\newblock {\em arXiv preprint arXiv:0805.4224\/} (2008).
\newblock


\bibitem{reschenhofer2015empirical}
{Thomas Reschenhofer} {and} {Florian Matthes}. 2015.
\newblock \showarticletitle{An empirical study on spreadsheet shortcomings from
  an information systems perspective}. In {\em International Conference on
  Business Information Systems}. Springer, 50--61.
\newblock


\bibitem{whittaker2003dynamics}
{Steve Whittaker}, {Loen Terveen}, {Will Hill}, {and} {Lynn Cherny}. 2003.
\newblock \showarticletitle{The dynamics of mass interaction}.
\newblock In {\em From Usenet to CoWebs}. Springer, 79--91.
\newblock


\end{thebibliography}
